\documentstyle[aps]{revtex}
\begin{document}

\title{Properties of two - dimensional dusty plasma clusters.}

\author{G.E.~Astrakharchik, A.I.~Belousov and Yu.E.~Lozovik\cite{*)}}

\address{Institute of Spectroscopy Russian Academy of Sciences,
142092 Troitsk, Moscow region, Russia}

\maketitle

\begin{abstract}
Two-dimensional classical cluster of particles interacting through a
screened Coulomb potential is studied.
This system can be used as a model for "dusty particles" in high-frequency
discharge plasma.
For systems consisting of $N = 2 - 40$ particles and confined by a harmonic
potential we find ground-state configurations, eigenfrequencies and eigenvectors
for the normal modes as a function of the Debye screening length $R_D$ in plasma.
Variations in $R_D$ cause changes in the ground-state structure
of clusters, each structural rearrangement can be considered as a phase
transition of first or second order (with respect to parameter $R_D$).
Monte Carlo and molecular dynamics are used to study in detail the melting of
the clusters as the temperature is increased.
By varying the density and the temperature of plasma, to which the
particles are immersed, one can modulate thermodynamical properties of
the system, transforming it in a controllable way to an ordered
(crystal-like), orientationaly disordered or totally disordered (liquid-like)
states.
The possibility of dynamical coexistence phenomena in small clusters
is discussed.
\end{abstract}


\section{Introduction}
\label{introduction}
Small charged particles of "dust" are rather common systems, and are
observed on different scales and in different environments:
clusters of dust in the interstellar medium, charged colloidal suspensions,
ordered structures in the gas discharge used in thermal
processing of materials are examples of such systems
\cite{Dp1,Nefedov,Khodataev,Studart}.
At present much attention is paid to experimental investigation of the
properties of "dusty plasma" which is a system
of small micrometer particles in a high frequency gas
discharge.
One of the main reasons of this attention paid to the artificial objects
like this one is the ability of a direct observation of their static and
dynamical properties.
The study of dusty plasma crystals and liquids 'in vitro' which is being
carried out in a number of laboratories around the world \cite{Nefedov} is of
a great importance for understanding the plasma properties and is a powerful
tool for the examination of melting, annealing and formation of defects of
different kinds.

Small particles immersed in a plasma may acquire large (up to $10^5 e$)
charges $Z e$ due to high mobility of plasma electrons.
The presence of plasma screening modifies the Coulomb interparticle
interaction and, with a good precision~\cite{inter1,inter2},
the system can be described as a system of particles interacting with
a Yukava-type pair potential.
Here we consider two-dimensional (2D) clusters of dusty particles
confined by an external harmonic potential of strength $\alpha$.
It is obvious that it is 2D system that realizes when one consider
particles of 'dust' immersed in a plasma discharge cloud
with the transverse dimension higher than the Debye screening length.
The energy of the system has the form:
\begin{eqnarray}
E = (Z e)^2 \sum\limits_{i<j}^N \frac{\exp{(-|{\bf r}_{ij}|/R_D)}}{|{\bf r}_{ij}|} +
\alpha \sum_{i=1}^N |{\bf r}_{i}|^2
\label{ham}
\end{eqnarray}
The Debye screening length in plasma is determined as
$R_D = \left( 4\pi q_i^2 n_i/k_bT_i + 4\pi e^2 n_e/k_bT_e\right)^{-1/2}$
where $q_i,n_i$ and $T_i$ are the charge, mean density and
temperature of plasma ions and $e,n_e,T_e$
are that of plasma electrons respectively.

The energy of a cluster, being written in dimensionless units
$r_0 = (Z e)^{2/3} / \alpha^{1/3}$ for distances and
$E_0 = \alpha r_0^2$ for energies becomes:
\begin{eqnarray}
E = \sum\limits_{i<j}^N \frac{\exp{(-\gamma |{\bf r}_{ij}|)}}{|{\bf r}_{ij}|^3}
+ \sum_{i=1}^N |{\bf r}_{i}|^2
\label{E}
\end{eqnarray}
where the dimensionless parameter $\gamma = r_0 / R_D$ defines the range of
the pair interaction potential.
From (\ref{E}) one can see that the thermodynamic state of a cluster
of a given number of particles is determined by two dimensionless parameters:
the inverse dimensionless screening length $\gamma$ and
the dimensionless temperature of "dusty" grains $\Theta = k_b T / E_0$.
The range of interaction between particles in a cluster $1/\gamma = r/R_D$
is controlled by the density and the temperature of a plasma (see above).

In this paper we consider the properties of two-dimensional (2D)
clusters (\ref{E}) as
a function of the number of particles $N<40$, the screening length $1/\gamma$
and the temperature $\Theta$.
We show (see Sec.~\ref{ground_config}) that
the change in the screening length (i.e. in parameter $\gamma$)
causes rearrangements of the ground-state structure of the cluster
at a set of points $\gamma^*$,
such structural transitions can be treated as phase transitions
of the first or the second order with respect to parameter $\gamma$.
In Sec.~\ref{transitions} we apply
molecular dynamics (MD) and Monte Carlo (MC) simulations
in a canonical ensemble in order to study the
thermodynamic properties of small clusters.
We show that in clusters of rather a small number of particles
and at a small enough plasma screening (at $\gamma < 10$),
as the temperature is increased,
the orientational disordering happens first, i.e
shells rotate with respect to each other by
losing their mutual orientation order.
At more higher temperatures a total disordering of cluster shells
takes place.

\section{Ground - state configurations}
\label{ground_config}

Ground-state configurations (see Table~1 and Fig.~1-3)
of the system (\ref{E}) have been found
with the help of the following methods:
1) The modified Newton method \cite{Bedanov};
2) The combination of "random search" and "gradient search" methods
\cite{dip_Bel}.
For the results to be more reliable, all configurations discussed below
have been independently obtained by both of this methods.
Of course, no one of the present methods of the minimization
of a multidimensional function is able to guarantee
that the configuration obtained is a global minimum one.
To overcome this difficulty we have used up to 200 randomly
distributed initial configurations.
This approach have also enabled us to investigate both the local minima
and their caption regions (i.e. "specific weights" of local minima).

At $\gamma \ll 1$ model (\ref{E})
describes the Coulomb cluster in a harmonic
trap, the system that have been actively studied both
experimentally\cite{trap1,trap2} and with the use of computer simulations
\cite{Mandel},\cite{Bedanov,Koulakov}.
In particular, the calculations carried out before have revealed
that particles in small finite systems arrange themselves into shells.
An analysis of
shell structures for different number of particles $N$
enables one to consider the system as belonging to some
period of a Mendeleev - type table.
This table can be viewed as a classical equivalent
to the well - known Periodic Table of elements.

The presence of parameter $\gamma$ which determines the range
of the pair potential enables one to
investigate the influence of this range on the structures and properties
of clusters.
The fact that cluster structure depends on the range of interaction
potential becomes obvious from Table~1 in which some ground-state
configurations for 2D clusters in a harmonic trap are presented.

As the value of parameter $\gamma$ changes,
rearrangements of ground-state structure take place,
each point $\gamma^*$ of any of these changes can be treated as a point
of phase transition of one kind or another.
Following by the approach used in work \cite{Partoens},
the order of these phase transitions can be determined from the
plot of the ground-state energy $E(\gamma)$,
the discontinuity in the $n-th$ derivative of $E(\gamma)$ with respect to
parameter $\gamma$ corresponds to the phase transition of $n-th$ order.
Another way to determine the order of the phase transition is to
analyse the behaviour of eigenfrequencies $\omega_i(\gamma), i=\overline{1,2N}$
for the normal modes: first order transition takes place at the point
$\gamma^*$ at which any of the eigenfrequencies exhibits a jump
while softening of any of the eigenfrequencies (when it becomes zero)
testifies about second order phase transition.

The eigenfrequencies of the cluster of $N=10$ particles
vs. screening parameter $\gamma \in [0,10]$ are presented in Fig.~1.
At points $\gamma \approx 1.4$ and $\gamma \approx 8.2$ of first order
transitions the eigenfrequencies exibit jumps that are clearly seen.
From Fig.~1b one can see that with a decrease in the interaction range
first, at $\gamma \approx 1.4$, the distribution of particles throughout
shells changes and the configuration typical for the Coulomb interaction
is replaced by one appropriate to the dipole cluster of 10 particles
($\{2,8\} \to \{3,7\}$).
Further reduction in the screening radius transforms (at $\gamma \approx 8.2$)
the cluster to the most "packed" state $\{2,8\}$ which is characteristic
of a system of hard spheres.

In Fig.~2 the ground-state energy of a cluster of $33$ particles are given.
At $\gamma \approx 3.751$ the first derivative of the
ground-state energy with respect to parameter $\gamma$
is discontinuous (see inset of Fig.~2a).
Investigation of cluster configurations shows that the numbers of particles
in two outer shells change here as $\{1,6,11,15\} \to \{1,6,12,14\}$.

The point of the first order phase transition $\gamma^*$ can be
determined as the point, where the energies of ground- and
the lowest excited (metastable) states are equal.
This statement is illustrated in Fig.~2b in which
both ground-state energy $E(\gamma)$ and energy $E^{(1)}(\gamma)$
of the lowest local minimum are plotted in the vicinity of the
transition point $\gamma^* \approx 3.751$.
From this figure one can see that the configuration of the global
minimum at $\gamma < \gamma^*$ corresponds to a local one at
$\gamma > \gamma^*$.

One can see that in the cluster of $37$ dipole particles
one of the particles is between the second and the third shell to form an
interstitial (analogous to the Frenkel defect in crystals) and to
make the division of the ground-state configuration into shells
ambiguous \cite{dip_Bel}.
The 'dusty' cluster of $37$ particles is supposed to exhibit a rich
variety of structural rearrangements while varying $\gamma$.
The investigation of this cluster at different values of screening strength
has revealed four phase transitions in the region
$\gamma \in [0,1.6]$ (see Fig.~3a), namely two second order transitions
(at $\gamma \approx 0.78$ and $\gamma \approx 1.22$)
and two first order transitions (at $\gamma \approx 0.52$ and
$\gamma \approx 1.34$).

From Fig.~3a on can see that the number of particles in
the outer shells changes at $\gamma \approx 0.52$.
It is worth while to note that a distinctive feature of the first
order phase transition is the abrupt change in the cluster structure,
Usually, this peculiarity exhibits as the change in the distribution
of particles throughout shells (like that one can observe
for the clusters of $10$ and $33$ particles, see Fig.~1,2).
This very change takes place at $\gamma = 0.52$ for the cluster involved.
However no apparent changes in structure of the cluster are seen at the
point $\gamma \approx 1.34$ of the first order phase transition.
More detailed study have shown that at this point
there exist a rotation of the third shell with respect to the fourth one.
This is illustrated in Fig.~3b which presents
the mutual orientational order parameter $g_{s_1 s_2}$
of different pairs of shells $\{s_1, s_2\}$  \cite{dip_Bel}, the value
which is very sensitive to the changes in mutual
orientation of cluster shells.

Subsequent increase in the parameter $\gamma$
leads to the realization of two other first order transitions which are
depicted in Fig.~3b.
In the first of them (at $\gamma \approx 7.015$) one of
the particles implants between the second and the third shells
(see Table~1 and the discussion above).
The corresponding transition can be written as
$\{1,7,13,16\} \to \{1,6,\overline{1},12,17\}$.
At $\gamma > 19$ the cluster becomes well - facetted and
has the most symmetrical structure \{1,6,12,18\}.
Note that in this region of $\gamma$
the minimal nonzero eigenfrequency $\omega_{min}$ corresponds to
twofold degenerate vibrations of the whole cluster in the harmonic trap
with the frequency $\omega_{min} = \sqrt{2}$.
Further decreasing in the range of the pair potential
does not lead to any structural rearrangements.

The study of Coulomb and dipole clusters have shown that
the basis for most configurations is provided by different parts
of 2D hexagonal lattice \cite{dip_Bel,Koulakov}.
When describing and analyzing the properties of such configurations
it is suitable to introduce into consideration
the "crystal shells" $Cr_c$ that are concentric groups of nodes of
ideal 2D crystal with $c$ nodes placed in the center of these groups.
Obviously, in view of the axial symmetry of the confinement potential,
we can concentrate on a finite number of the most symmetrical crystal
shells which, by the number of particles in the center of the system,
can be divided into the following groups:
$Cr_{1}$, $Cr_{2}$, $Cr_{3}$, $Cr_{4}$.
With the help of the crystal shell concept we have found
that changes in the ground state structure of "dusty" clusters,
as parameter $\gamma$ is increased,
comes in such a way, as to fill the maximal number of crystal shells.

\section{Phase transitions}
\label{transitions}

One of the distinctive peculiarities of small clusters is
the existence of two stages of their disordering \cite{Mandel,Rakoch_log}:
an intershell (orientational melting of shells $s_1$ and $s_2$ at the
temperature $\Theta_{s_1 s_2}$) and a radial disordering
(total melting at temperature $\Theta_f$).
The analysis of eigenfrequencies shows that the clusters with the small
values of lowest nonzero eigenfrequencies $\omega_{min}$ have the
eigenvectors corresponding to mutual rotations of cluster shells.
Such clusters have low temperatures
$\Theta_{s_1 s_2}$ of intershell disordering.
It is obvious that changes in the cluster structure caused by
variations in the control parameter $\gamma$ lead to the modulation of the
temperatures of both orientational $\Theta_{s_1 s_2}$ and total $\Theta_f$ disordering.
Moreover, the phenomenon of the orientational disordering
may disappear at all if the cluster has a well packed structure.
The results of our simulations have proved this suggestion.

The dependencies of the mutual orientational order parameter
$g_{2 1}(\Theta)$ for two-shell cluster of $N=10$ particles
at several values of parameter $\gamma$ are given in Fig.~4a.
It is evident that $g_{s_1 s_2}$ drops to zero at the point of relative
disordering (mutual rotation of shells $s_1$ and $s_2$) \cite{notation}.
One can see from Fig.~4a that the change in the system configuration
$\{2,8\} \to \{3,7\}$ which occurs at $\gamma \approx 1.4$ (see Fig.~1)
leads to a sharp decrease in the orientational disordering temperature:
$\Theta_{21}(\gamma < 1.4) \approx 1.3 \cdot 10^{-4} \to
\Theta_{21}(\gamma > 1.4) \approx 0.7 \cdot 10^{-5}$.

The cluster is well - packed in the region $\gamma > 8.2$ (see Fig.~1) and
that is why it does not experience the orientational melting, when
an increase in the temperature leads directly to the interchange
of particles between shells at $\Theta \approx 10^{-3}$.
This can be seen from the analysis
of radial square deviations $u^2_r$:
\begin{eqnarray}
u_{r}^2 = \frac{1}{N} \sum\limits_{i}
\left[ \left< |{\bf r}_i |^2 \right> - \left< |{\bf r}_i | \right>^2 \right]
\label{u_r}
\end{eqnarray}
The dependence $u^2_r(\Theta)$ is given in Fig.~4b,
also shown are analogous curves at $\gamma = 1$ and $\gamma = 2$.
One can see that even the slightest variation in the value of control
parameter $\gamma$ may change the temperature of the total melting
up to orders.

The changes in the interaction potential lead to the modification
of the structure of the energy surface which determines the type and
the distinctive features of the phase transitions.
For this reason, one can suppose that at some values of the parameter
$\gamma$ the system can have very interesting thermodynamic properties.
In Fig.~5a the dependence of the radial square deviations (\ref{u_r}) of
four - shell cluster of $N=33$ particles at $\gamma = 3.76$ is given.
The graph has a number of plateaus located in different temperature intervals.
A detailed investigation has shown that
the regions of a sharp increase in $u_{r}^2$
correspond to the radial disordering of different pairs of shells:
particles start to interchange between the third and the fourth shell
at temperature $\Theta^f_{3 4} \approx 5 \cdot 10^{-4}$ and
between the second and the third -- at $\Theta^f_{2 3} \approx 0.005$.
The total melting of the
cluster takes place at $\Theta^f \approx 0.01$.

Some useful information about the character of the disordering
considered can be obtained by exploring the local minima
distribution $\rho(E_{loc})$.
In order to estimate this histogram, at each measurement time point
we have performed several hundreds of gradient search iterations
to find the nearest local minimum with the energy $E_{loc}$.

In Fig.~5b the local minima distribution of the system of 33 particles
at $\gamma = 3.76, \; \Theta = 10^{-4}, \; \Theta = 8 \cdot 10^{-3}$
is shown.
In an entirely ordered state (at $\Theta = 10^{-4}$) the system lives
in the vicinity of the global minimum (with energy
$E = 64.795946$ and the structure $\{1,6,12,14\}$).
At $\Theta = 8 \cdot 10^{-3}$ the cluster can be also found at the lowest local
minimum $E^{(1)} = 64.795975$ with the configuration $\{1,6,11,15\}$
(see Fig.~2b).

Considering the results stated above one can conclude that the first
disordering seen in the temperature interval
$\Theta \in [10^{-4}, 10^{-3}]$ (see Fig.~2)
corresponds to the nonzero probability of the cluster to be found in
the state $\{1,6,11,15\}$ which is metastable at the
given value of the parameter $\gamma$.
Such changes in the distribution of particles throughout shells demand
the overcoming of potential barrier that, knowing about
large specific weights of both "ground" and "excited" states, allows one to
treat this temperature interval as that of the dynamical coexistence
of two cluster forms $\{1,6,12,14\} \rightleftharpoons \{1,6,11,15\}$.

\section{Conclusions}
\label{conclusions}

In this letter we have presented the results of a study
of finite "dusty plasma" particle system.
As a function of Debye screening length $R_D$ (for the particle charge
in the plasma) we have found ground-state
configurations of clusters consisting of $N \le 40$ particles, their normal
modes eigenvalues and the corresponding eigenvectors.
The clusters undergo structural transitions
which manifest itself as phase
transitions of first or second orders with respect to parameter $R_D$.
At points of first order transitions cluster coordinates
experience jumps which lead either to the change in the
shell distribution or to the rotation of some shells relative to each other.
At points of second order
transitions one of the eigenfrequencies softens and particle coordinates
change continuously.

By varying $R_D$ (for example, by varying the temperature and
the density of the plasma)
one can modulate thermodynamic properties of the system
and considerably change the temperatures of both the
orientational and the total disordering.
It have turned out that for some clusters, as
the screening becomes sufficiently high,
the disappearance of the orientational melting of different parts
of the system takes place and an increase in the temperature
leads staightway to the interchanges of particles between shells.

\vspace{1cm}

{\bf Acknowledgments}.
This work was partially supported by Russian Fund of Basic Researches,
INTAS and the program "Physics of solid state nanostructures"


Table~1. \\
Ground-state shell structures $\{N_1,N_2,...\}$ for dipole, Coulomb and
logarithmical clusters of $N$ particles confined in a harmonical potential.

\vspace{0.5cm}

\noindent
{\small
\begin{tabular}{|p{1cm}|p{4cm}|p{4cm}|p{4cm}|}
\hline
$N$ & Dipole cluster & Coulomb cluster & Logarithmical cluster\\
\hline
9  & 2,7         & 2,7       & 1,8\\
10 & 3,7         & 2,8       & 2,8\\
11 & 3,8         & 3,8       & 3,8\\
...&...          &...        &...\\
32 & 1,6,12,13   & 1,5,11,15 & 4,11,17\\
33 & 1,6,12,14   & 1,6,11,15 & 5,11,17\\
34 & 1,6,12,15   & 1,6,12,15 & 1,5,11,17\\
...&...          &...        &...\\
36 & 1,6,12,17   & 1,6,12,17 & 1,6,12,17\\
37 & 1,6,1,13,16 & 1,7,12,17 & 1,6,12,18\\
38 & 2,8,13,15   & 1,7,13,17 & 1,6,12,19\\
\hline
\end{tabular}
}

\vspace{2cm}

Fig.~1. \\
Eigenfrequencies (a) and the lowest nonzero eigenfrequency
$\omega_{min}$ (b) of the 'dusty' cluster of $10$ particles.
Inset: ground state configurations in three different regions of
control parameter $\gamma$.

\vspace{1 cm}

Fig.~2. \\
Cluster of $33$ particles.
a) The first derivation of the cluster ground energy with respect to $\gamma$.
In the inset the region of the first order phase transition is shown
on an enlarge scale.
b) Energies and configurations of the lowest local minimum
(with the energy $E^{(1)}$ measured from the ground state energy $E$)
of the cluster in the region of the phase transition.

\vspace{1 cm}

Fig.~3. \\
Cluster of $37$ particles.
 ) The lowest nonzero eigenfrequency $\omega_{min}(\gamma)$ and
mutual orientational order parameter of different pairs of shells
$g_{s_1 s_2}(\gamma)$.
b) The regime of strong plasma screening.
Eigenfrequency $\omega_{min}$ corresponds at $\gamma > 19$
to the motion of the cluster at a whole in the confinement.

\vspace{1 cm}

Fig.~4. \\
Two - shell cluster of $10$ particles.
 ) Thermodynamical mean of the mutual orientational order parameter
$<g_{21}>(\Theta)$ at different values of $\gamma$.
b) radial mean - square deviations od particles vs. temperature
$u^2_r(\Theta)$.

\vspace{1 cm}

Fig.~5. \\
Four - shell cluster of $33$ particles
a) radial mean - square deviations.
b) local minima distribution histogramm $\rho(E^{(loc)})$ of the cluster
in the ordered state (at $\Theta = 10^{-4}$)
and at $\Theta = 8 \cdot 10^{-3}$, when there is an interchange of particles
between the third and the fourth shells.


\begin{references}

\bibitem{*)} e-mail: lozovik@isan.troitsk.ru

\bibitem{Dp1}      C.H.~Chiang and L.~I,
                   Phys. Rev. Lett. {\bf 77}, 646, (1996).

\bibitem{Nefedov}  A.P.~Nefedov, O.F.~Petrov, V.E.~Fortov,
                   Uspekhi Fiz. Nauk (in Russian) {\bf 167}, 1215, (1997).
                   V.E.~Fortov, A.P.~Nefedov, O.S.~Vaulina et. al.,
                   JETP {\bf 114}, 2004 (1998).

\bibitem{Khodataev} Y.K.~Khodataev, S.A.~Krapak, A.P.~Nefedov, O.F.~Petrov,
                    Phys. Rev. E {\bf 57}, 7086 (1998)

\bibitem{Studart} L.~Cardido, J.P.~Rino, N.~Studart, F.M.~Peeters,
                   J. Phys.: Cond. matt.  {\bf 10}, 11627 (1998).

\bibitem{inter1}    D.P.~Resendes, J.T.~Mendonca, P.K.~Shukla,
                    Phys. Lett. A {\bf 239}, 181 (1998);
                    M.~Namby, S.V.~Vladimirov, P.K.~Shukla,
                    Phys. Lett. A {\bf 203}, 40 (1995).

\bibitem{inter2}    A.~Melzer, V.A.~Shveigert, I.V.~Sweigert, A.~Homann,
                    S.~Peters, and A.~Piel,
                    Phys. Rev. E {\bf 54}, R46 (1996).

\bibitem{Bedanov} V.M.~Bedanov and F.M.~Peeters
                  Phys. Rev. B. {\bf 49}, 2662 (1994);
                  V.~Shweigert and F.M.~Peeters
                  Phys. Rev. B. {\bf 51}, 7700 (1995);
                  I.V.~Shweigert, V.A.~Shweigert and F.M.~Peeters
                  Phys. Rev. B. {\bf 54}, 10827 (1996);

\bibitem{dip_Bel} A.I.~Belousov and Yu.E.~Lozovik, {\it Conference
                  on Computational Physics "CCP-1998"};
                  preprint cond-mat/9803300.

\bibitem{trap1}  R.~Bl\"umel, J.M.~Chen, E.~Peik, et. al.
                 Nature {\bf 334}, 309 (1988).

\bibitem{trap2}  M.~Drewsen, C.~Brodensen, L.~Hornekar and J.S.~Hangst,
                 Phys. Rev. Lett. {\bf 81}, 2878 (1998).

\bibitem{Mandel} Yu.E.~Lozovik, UFN (Uspechi Fiz. Nauk, in russian)
                 {\bf 153}, 356 (1987);
                 Yu.E.~Lozovik, V.A.~Mandelstam,
                 Phys. Lett. A {\bf 145}, 269 (1990);
                 Phys. Lett. A {\bf 165}, 469 (1992);

\bibitem{Koulakov} A.A.~Koulakov and  B.I.~Shklovskii,
                   Phys. Rev. B {\bf 57}, 2352 (1998).

\bibitem{Rakoch_log} Yu.E.~Lozovik, E.A.~Rakoch,
                     Phys. Rev. B {\bf 57}, 1214 (1998).

\bibitem{Partoens} B.~Partoens, V.A.~Shweigert and F.M.~Peeters
                   Phys. Rev. Lett. {\bf 79}, 3990 (1997).

\bibitem{notation} the quantities $\psi_s$ and
$g_{s_1 s_2}=\psi_{s_1} \psi_{s_2}^*$ are analogous to
the parameter $\psi_6$ and the correlation function $g_6(r)$
of infinite 2D systems, the softening of the correlation function
$g_6(r) \to 0,\quad r \to \infty$ corresponds to the relative orientational
disordering of remote parts of the system
(with the absence of the translational order).

\end{references}
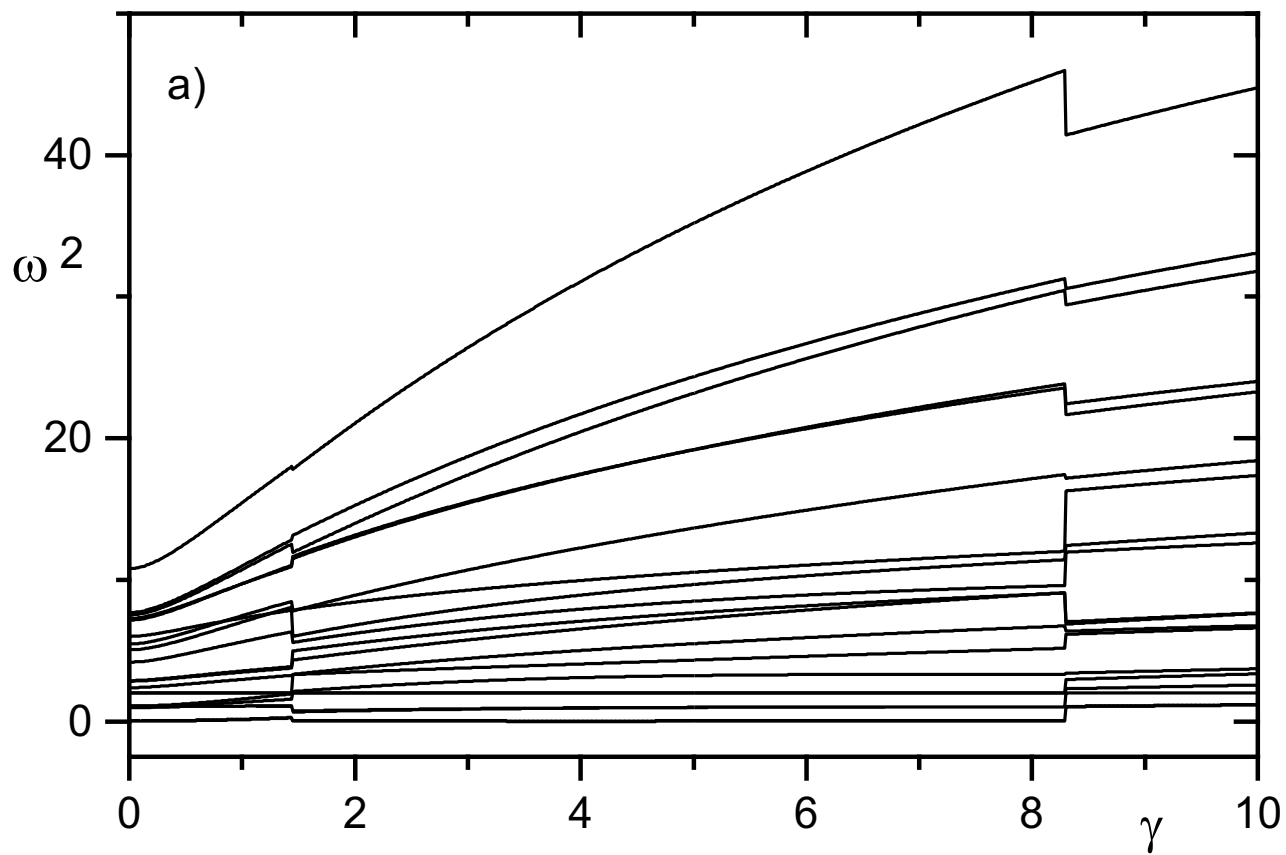
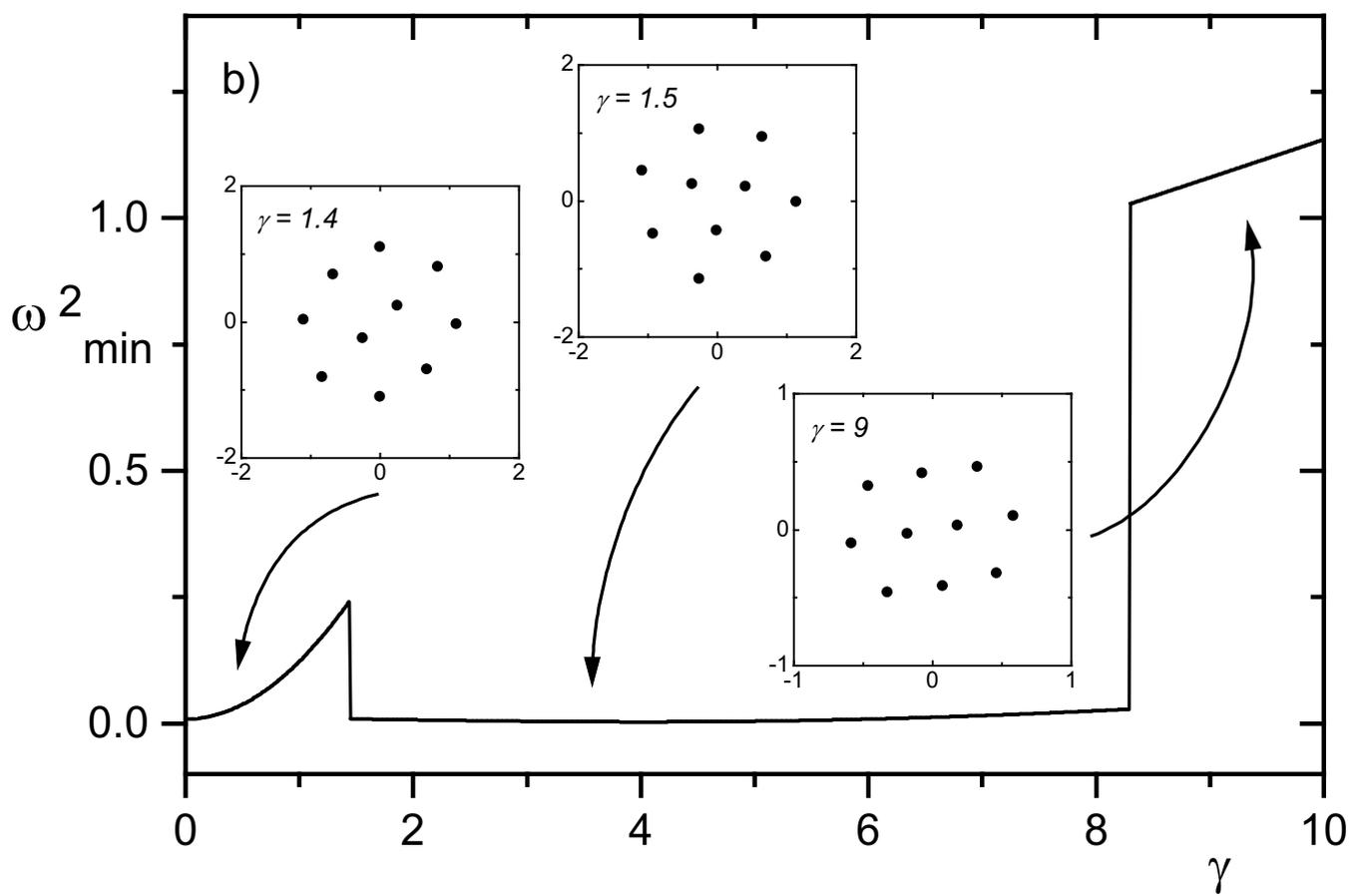
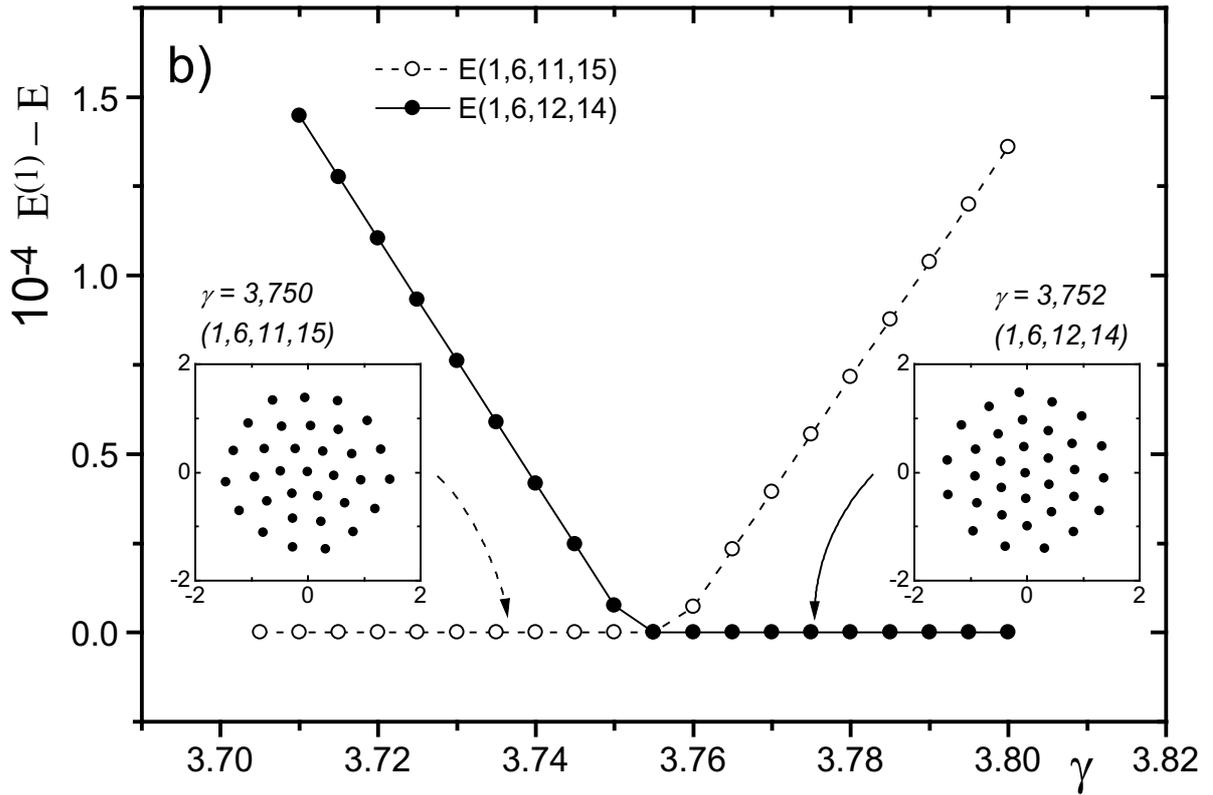
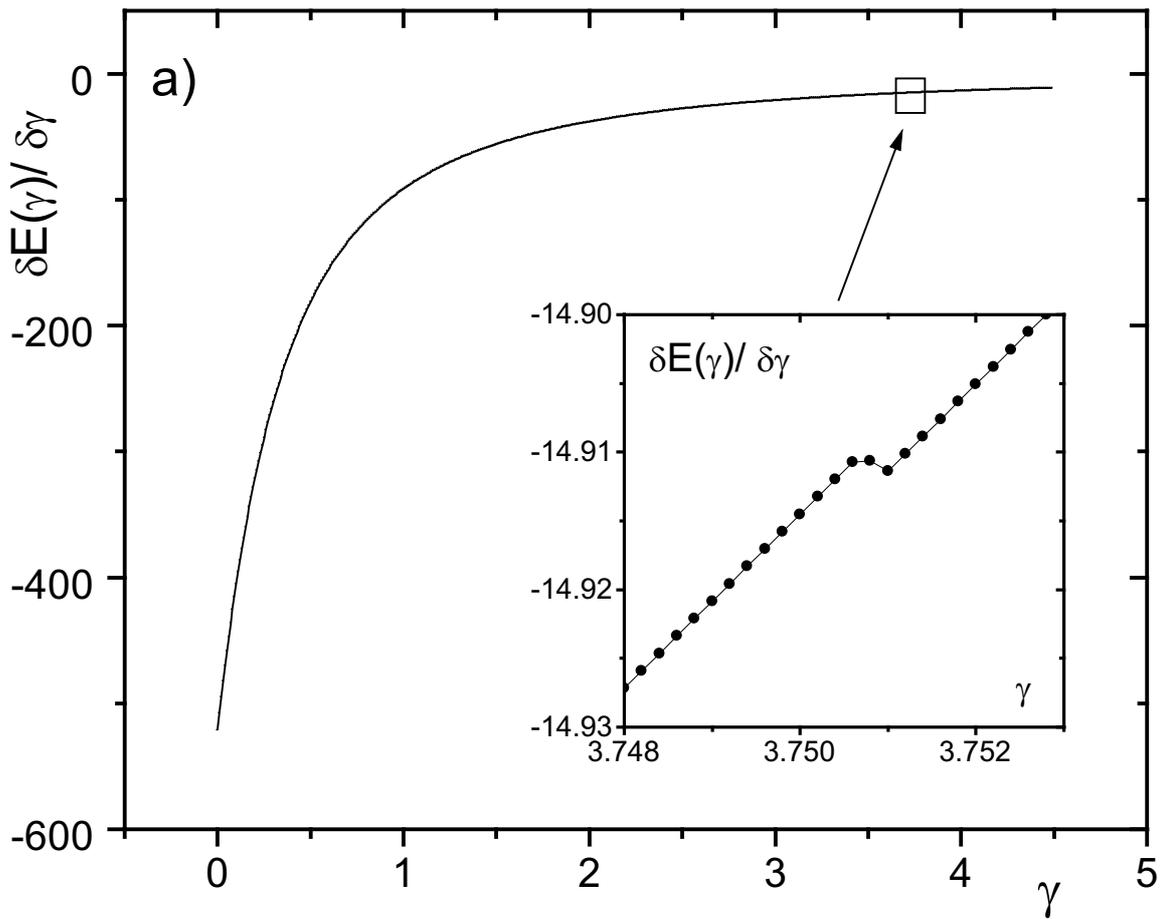
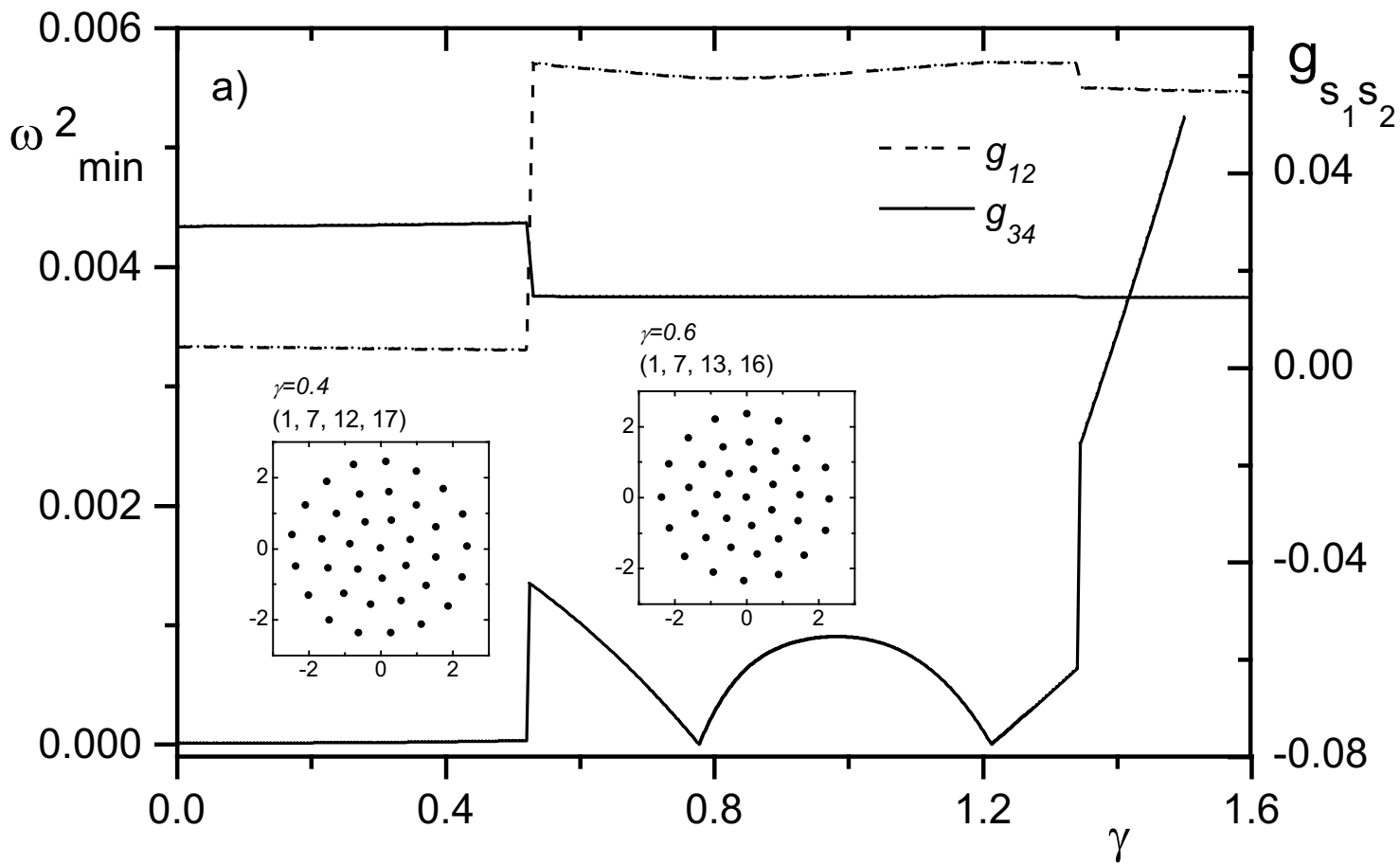
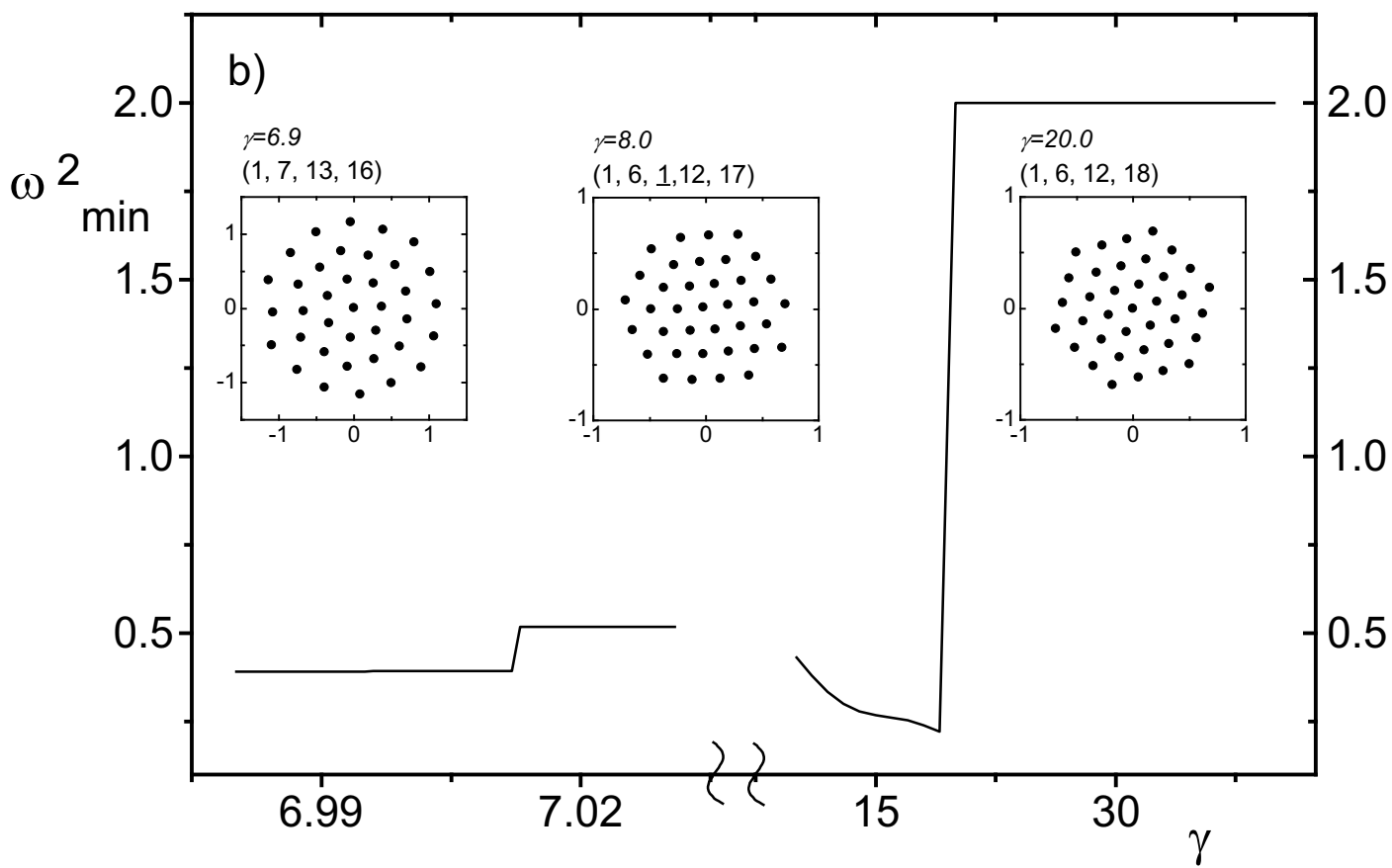
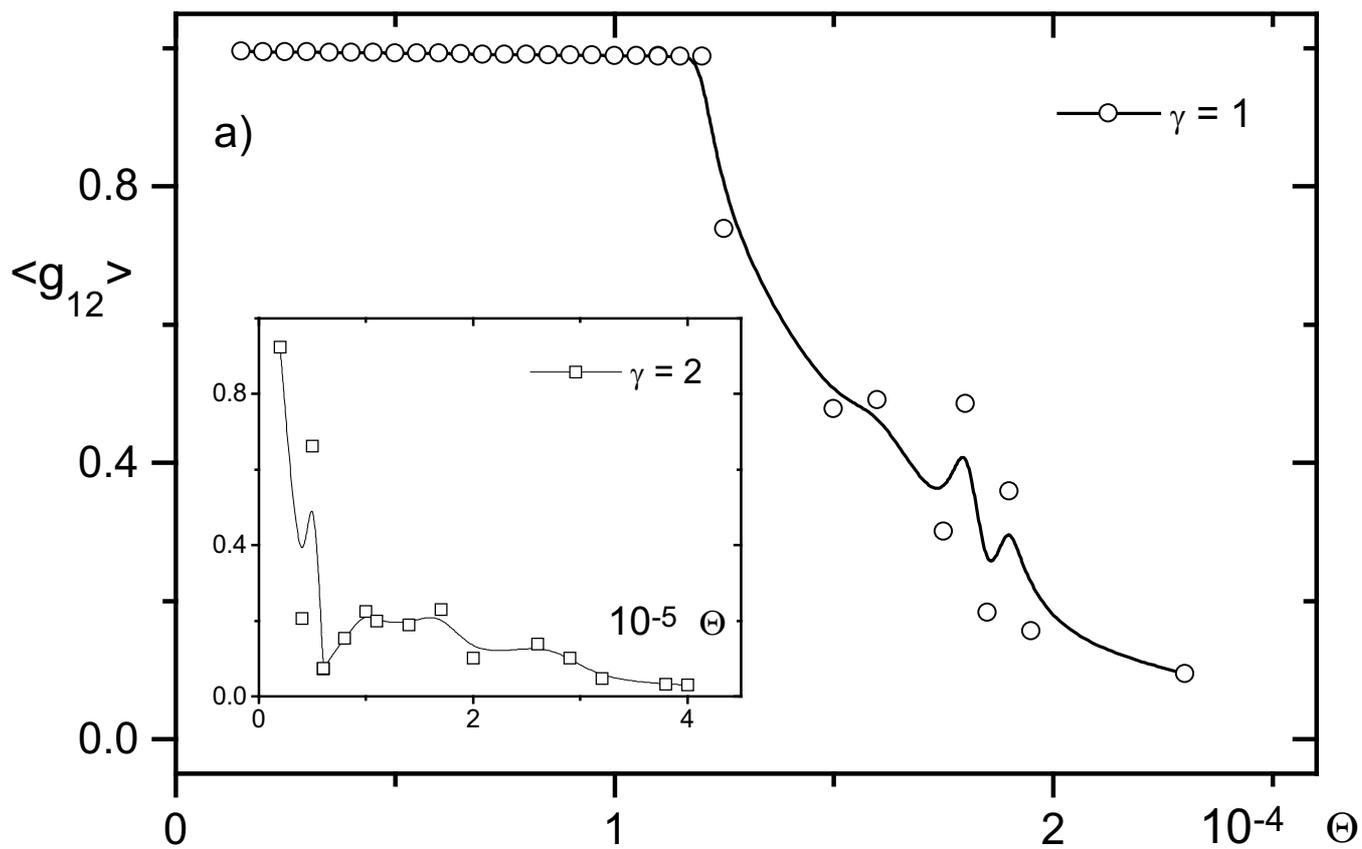
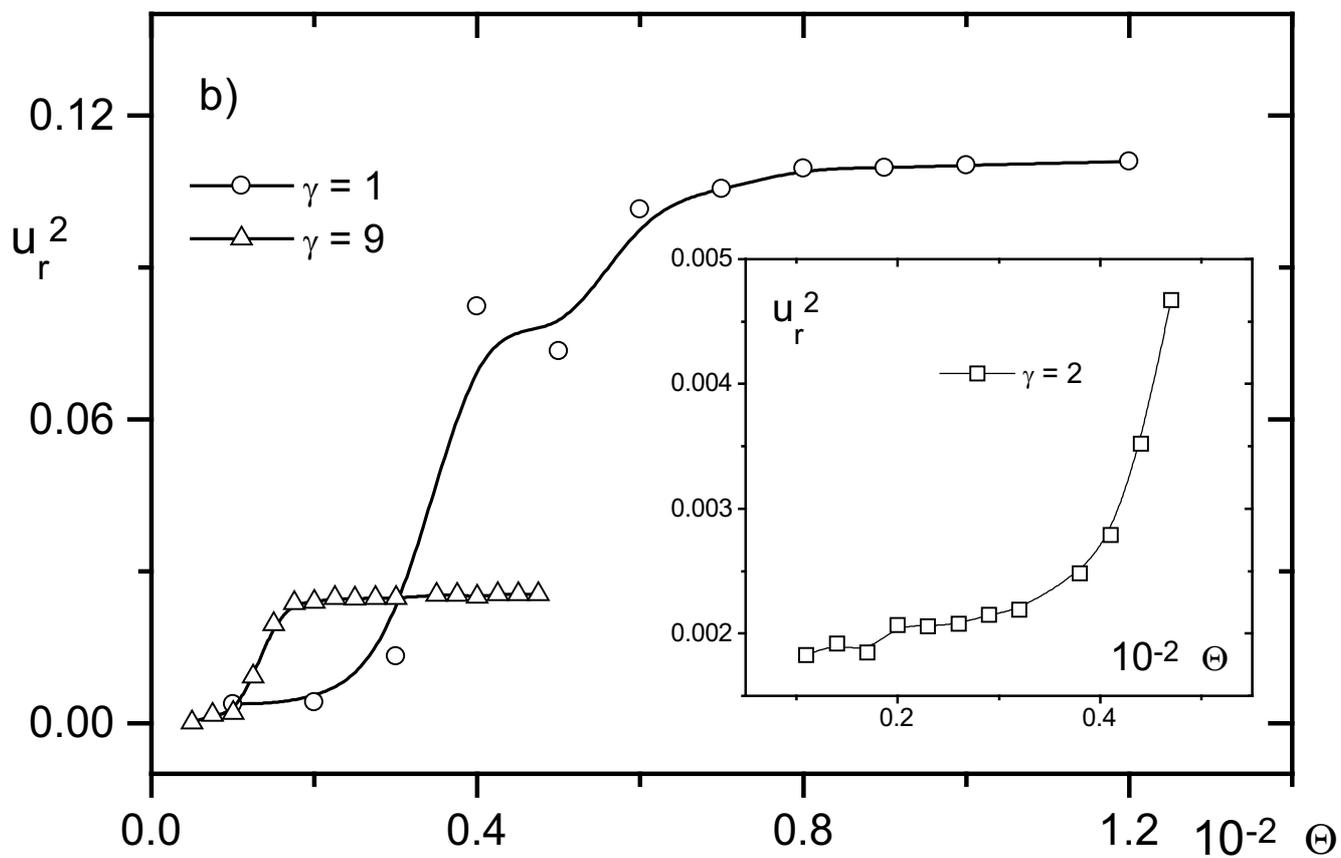
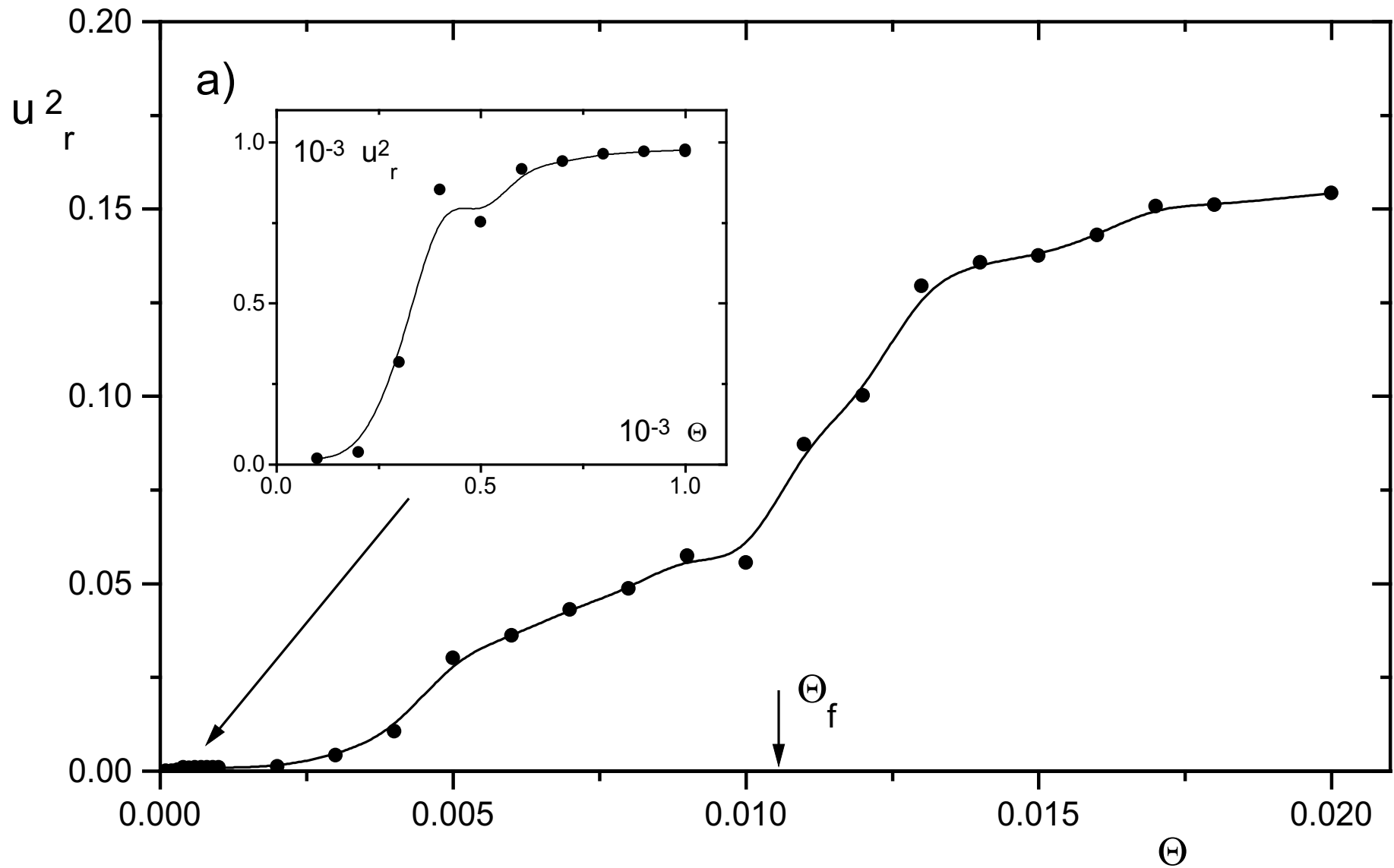
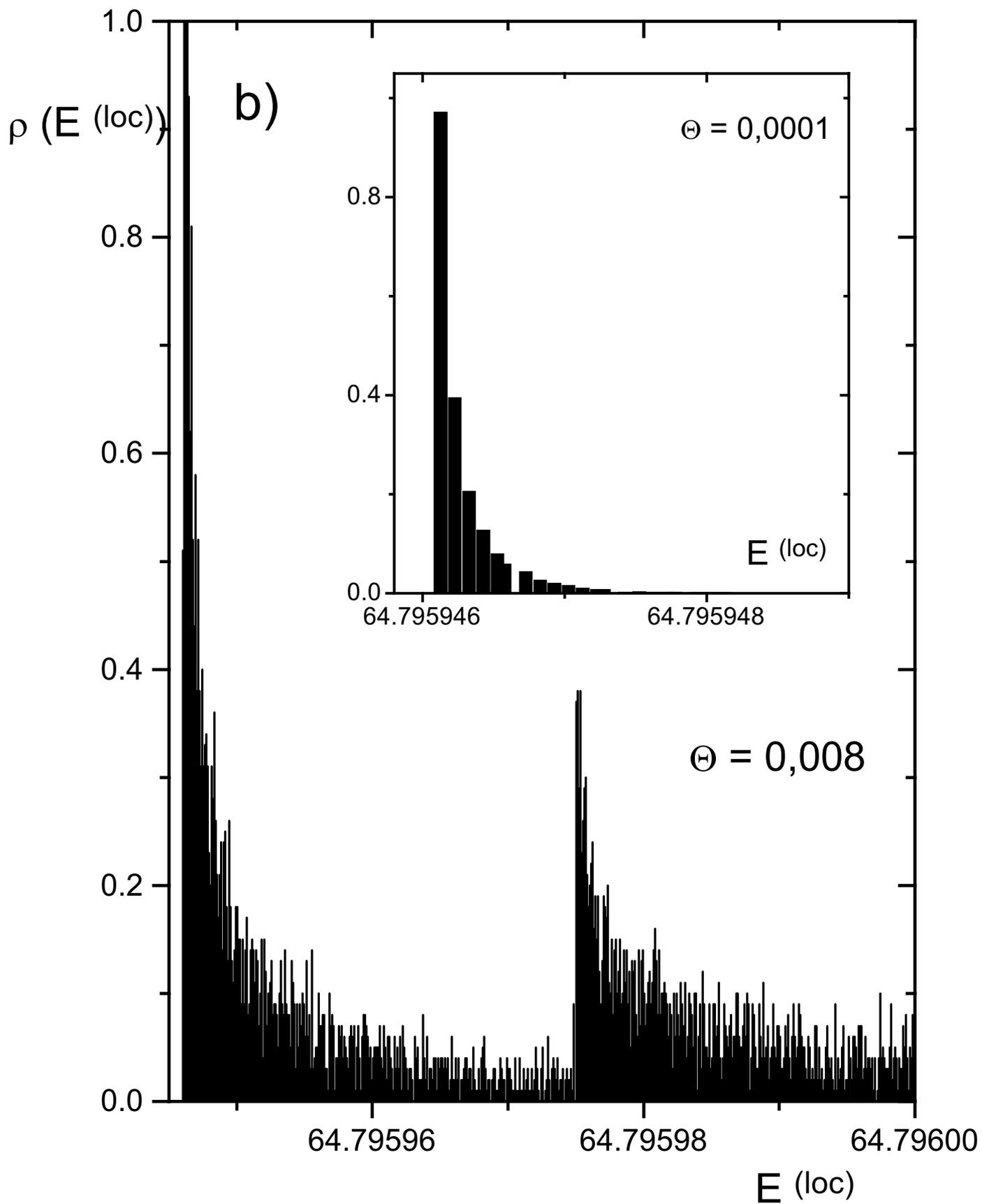
\end{document}